\documentclass[12pt,fleqn]{article}
\usepackage{graphicx}
\usepackage[cp1251]{inputenc}
\def\be{\begin{equation}}
\def\ee{\end{equation}}
\def\bea{\begin{eqnarray}}
\def\eea{\end{eqnarray}}
\def\beq{\begin{eqnarray*}}
\def\eeq{\end{eqnarray*}}

\def\ba{\begin{array}}
\def\ea{\end{array}}

\def\cor{{\rm cor}}

\begin{document}
\begin{center}
 {\large\bf Stochastic Acceleration in Strong Random Fields}

\vspace{0.3cm} {\bf A. Zagorodny$^a$, V. Zasenko$^a$, J. Weiland$^b$ }

$^a${\it Bogolyubov Institute for Theoretical Physics, 03143 Kiev,
Ukraine}


$^b${\it Department of Electromagnetics, Chalmers University of
Technology and \\ Euroatom-VR Association, 41296 G\"oteborg,
Sweden}

\end{center}

\begin{abstract}
Diffusion of particles in velocity space undergoing turbulent field was extensively studied in the problem of
warm beam relaxation. Under low field intensities the diffusion is described by the Fokker-Planck equation with
the diffusion coefficient given by quasilinear theory. This diffusion coefficient is calculated on the free
particle propagator and for weak fields its renormalization due to orbit diffusion is not necessary. To study
effects which should be taken into account when the intensity of the turbulent field is increased a numerical
simulation of particle motion in the external field of Langmuir waves with given $k$-spectrum and random phases
is done. For strong fields meansquare velocity evolution shows that ballistic regime in the very beginning is
changed for oscillatory one in the intermediate stage and later for diffusion. Asymptotically it behaves like
fractional power of elapsed time with the exponent dependent on the particular field spectrum. Such evolution in
the whole temporal interval of simulation is recovered from the numerical solution of generalized Fokker-Planck
equation with time dependent diffusion coefficient obtained from the microscopic approach. The analytical
approximation for this solution is also given.
\end{abstract}

\section{Introduction}
Diffusion of particles in random external fields could be considered in relation to a general problem of
transport in plasmas. The test particle approach, which is simpler and more controllable than selfconsistent
one, helps to analyze some particular aspects of turbulent transport.

One of not clarified issues reported in papers [1,2] concerns the enhancement of a diffusion coefficient in
velocity space as compared to its quasilinear value. It was supposed [1] that the enhancement was caused by
peaks in spectrum being formed in a course of evolution. The  tendency of formation of nonuniform structures in
plasma-beam system  was pointed out in the works [3]. Here we are interested in what effects are to be taken
into account when the spectrum of external field become stronger and narrower, i.e. more peaked.

We have made a direct simulation of test particle motion in prescribed random fields for a variety of spectra,
and found from obtained data the evolution of average velocity and dispersion. Then the generalization of the
Fokker-Planck equation were considered in order the solutions would be consistent with simulation in wide range
of variation of external field spectra.

As it was expected for low intensity and broad spectrum (small
Kubo number) solutions of Fokker-Planck equation with quasilinear
diffusion coefficient agree with the results of simulation. When
the form of spectrum is taken peaked (Kubo number becomes larger than the
unit) the velocity dispersion grows very fast at a
time less than the field correlation time, and such jump of dispersion
on the very early (prekinetic) stage could give a substantial
contribution to overall dispersion. In this case the distribution
function is governed by a Fokker-Planck equation with time dependent
diffusion coefficient. Solutions of the Fokker-Planck equation with
quasilinear and time dependent diffusion coefficients were found
numerically, and an analytical approximation for them was proposed
as well.

The numerical experiment gives the power law behavior of the dispersion at simulation times, and solutions of
the Fokker-Planck equation shows it for extended interval. However the exponent is not unique for all spectra of
the same type, as this follows from scaling consideration [4], but depends on the particular spectrum.

\section{Numerical model}

We consider the motion of noninteracting particles in an external random electric field. The potential of the
field is taken as a superposition of $M$ waves [5] \be \varphi(x,t)=\sum_{i=1}^M\, \varphi_i\cos \left(\omega
t-k_ix+\alpha_i\right) \ee with fixed frequency $\omega$ and wave numbers from the interval $(k_0-2.5\Delta k,\,
k_0+2.5\Delta k)$. The  total intensity of the field $\varphi_0^2$ is distributed between the partial waves
according to the Gaussian \be \varphi_i^2={2\over \sqrt{\pi}} \varphi_0^2\, {\delta k\over \Delta k}
\exp{-\left({k_i-k_0\over \Delta k}\right)^2}, \ee where $k_0$ is the central wave number, $\Delta k$ is the
width of the spectrum, $\delta k=k_{i+1}-k_i=5\Delta k/M$.

Each realization of the field (1) is characterized by unique set
of random phases $\{\alpha_i\}$. For a given realization of the
potential (1) the  equations of particle motion,
\bea
\dot{x} &\!\!\!=\!\!\!& v,\nonumber \\
\dot{v} &\!\!\!=\!\!\!& {e\over m} E(x,t),\qquad E(x,t)=-\frac{\partial }{\partial x}
\varphi(x,t),  \eea with the initial conditions $x(0)=0$, $v(0)=v_0$ are integrated
numerically to obtain the particle trajectory in velocity space $v(t)$. Average particle
velocity $\bar{v}(t)$ and velocity dispersion $\langle \Delta v^2\rangle_t$ are found by
averaging over $N$ realization
\bea \bar{v}(t) &\!\!\!=\!\!\!& {1\over N} \sum_{j=1}^N\, v_j(t), \nonumber
\\
\langle \Delta v^2\rangle_t  &\!\!\!=\!\!\!& {1\over N}
\sum_{j=1}^N\, \left( v_j(t)-\bar{v}(t)\right)^2.
\eea

In simulations length is normalized to $2\pi/k_0$, time to $2\pi/\omega$, dimensionless potential and spectrum
width are \beq \sigma={e\over m}\varphi{k_0^2\over\omega^2} \qquad \mbox{and}\qquad d={\Delta k\over k_0}. \eeq
Kubo number, $Q$, which is the ratio of the correlation time to characteristic period of particle oscillations
for this model is \be Q=\sqrt{\sigma}/d. \ee Overlap parameter \beq A_j=4\pi^2{e\over m} \varphi_j {k_j^4\over
\delta k^2\omega^2} \eeq much exceeds the unit for most harmonics $j$ except of few at the wings of Gaussian
distribution (2), and particle motion can be treated as stochastic. Note, that the random phase ensemble we used
here does not provide stochastization by itself, but gives the explicit way for calculation of Euler correlation
function of fields. According to Eqs.~(1), (2) the correlation function for the potential $\langle
\varphi^2\rangle_{xt}$ is of the form \be \langle \varphi^2\rangle_{xt} =\varphi_0^2 \exp-{\left(\Delta k\,
x\right)^2\over4} \cos (\omega t-k_0x). \ee \noindent Obtained in simulation $\bar{v}(t)$ and  $\langle \Delta
v^2\rangle_t$ will be compared in Section 4 with numerical and approximate analytical solutions of the
Fokker-Planck equation.

\section{Equation for distribution function and approximate analytical solution }

Introduce here the particle distribution function $f(v,t)$ as microscopic distribution function averaged over
random phase ensemble and integrated over the spatial variable $x$. As far as the averaging over the ensemble of
random phase does not imply the averaging over any small but finite time scale, the distribution function is
defined at all time scales, as well for $t<\tau_\cor$, i.e. in  prekinetic stage. Generalized  Fokker-Planck
equation for $f(v,t)$ in external fields could be obtained from Ref.\,6 in the form

\be \frac{\partial f}{\partial t} =\frac{\partial }{\partial v} D(v,t) \frac{\partial }{\partial v} f(v,t) \ee
with a time dependent diffusion coefficient
\bea D(v,t)=\left({e\over m}\right)^2 \int\limits_0^t \langle
E^2\rangle_{v\tau,\tau} d\tau. \eea

For correlation function (6) it takes the form \bea D(v,t) = {1\over 2}\left({e\over m} \varphi_0 \Delta
k\right)^2 \int\limits_0^t d\tau \exp-({1\over2}\Delta kv\tau)^2\nonumber
\qquad \qquad \qquad \qquad \qquad \qquad \qquad\\
\times\bigg\{\left(1+2({k_0\over \Delta k})^2 -{1\over2}(\Delta k v \tau)^2\right) \cos (\omega-k_0 v)\tau
+2k_0v\tau \sin (\omega-k_0 v)\tau \bigg\}. \eea

It will be shown that for moderate Kubo number the agreement with simulation is recovered by more accurate
treatment of distribution function evolution on early stage $t<\tau_\cor$. Here the use of a time dependent
diffusion coefficient is required.

The asymptotic value of $D(v,t)$ at large times $t \gg \tau_\cor$ gives the well known quasilinear diffusion
coefficient $D_{ql}(v)$ \be D_{ql}(v)=D(v,t\rightarrow\infty). \ee When the correlation function is given by Eq.
(6) the quasilinear diffusion coefficient takes the form \be D_{ql}(v) =\left({e\over m} \varphi_0 \right)^2
\sqrt{\pi} {\omega^2\over \Delta k |v|^3} \exp -({\omega-k_0 v\over\Delta k\, v})^2 . \ee In the following
section it will be shown that in the cases of narrow and/or high intensive spectrum it is important to retain
the dependence of diffusion coefficient on time.

Eqs.~(7), (9) or (7), (11) which determine the evolution of $f(v,t)$ with time dependent or, respectively,
quasilinear diffusion coefficient are solved numerically. Initial condition for $f(v,t=0)$ was taken as a narrow
Gaussian distribution that approximate $\delta (v-v_0)$. Then average velocity and dispersion are calculated as
\be \bar{v}(t)= \int dv\, v\, f(v,t), \ee \be \langle\Delta v^2\rangle_t=\int dv(v-\bar{v}(t))^2 f(v,t). \ee
They are compared in the following section with $\bar{v}(t)$ and $\langle\Delta v^2\rangle_t$ obtained in
simulation.

In Fig.~1 the time dependent diffusion coefficient $D(v,t)$
for narrow spectrum of external field
along with its profiles are shown.

\begin{figure}[h]
\includegraphics[width=7.5cm]{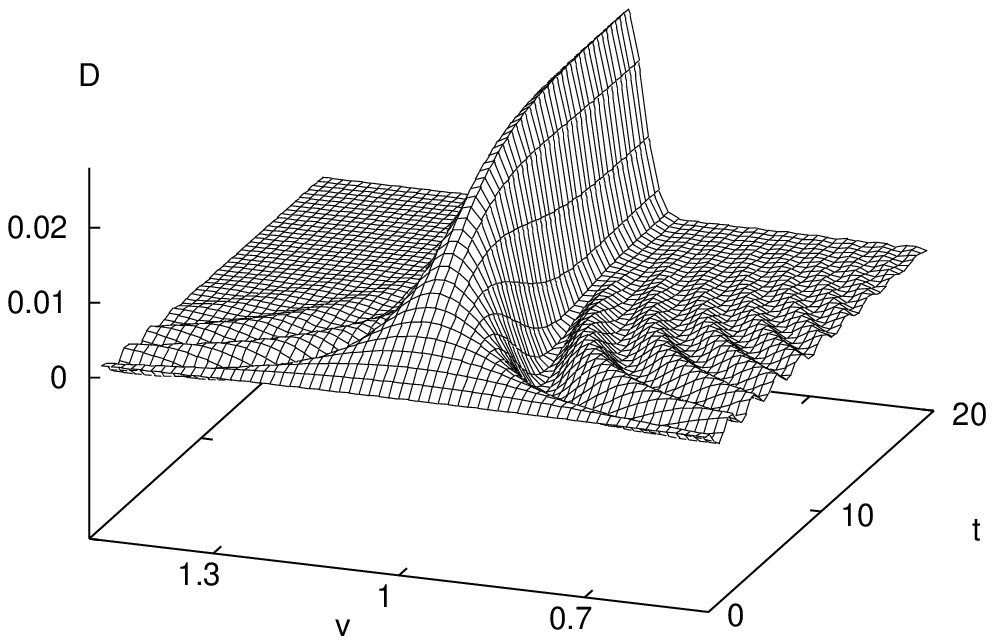}
\includegraphics[width=7.5cm]{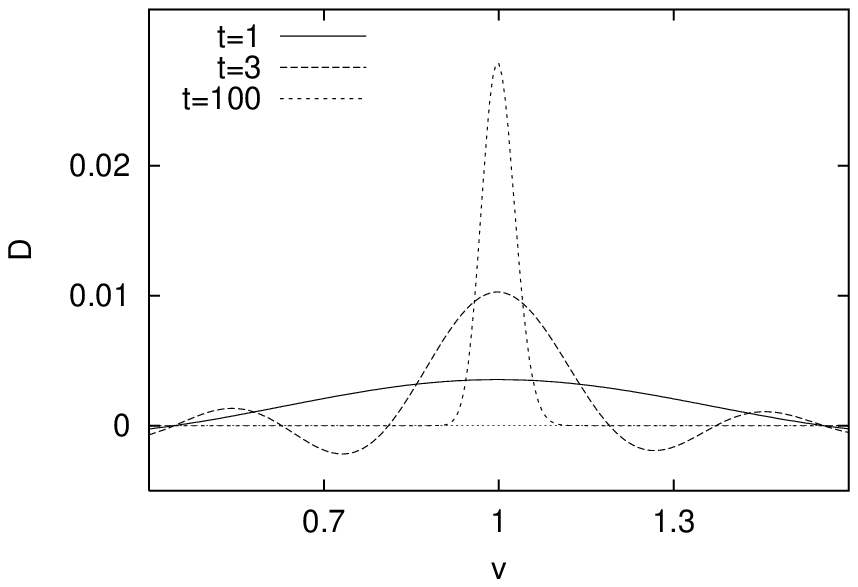}\\
\begin{center}
{\it Fig.\,1. Time dependent diffusion coefficient $D(v,t)$ for times less than the correlation time $\tau_\cor$ ({\it
left}), sections of $D(v,t)$ at $t=1$, $3$ and $100$ ({\it right}),\quad $\sigma=0.01$, $d=0.04$.}
\end{center}
\end {figure}

\vspace{0.3cm} \noindent Time dependent diffusion coefficient (9) evolve from very broad distribution through
oscillating regime to its asymptotic quasilinear value (11).

Approximate WKB-type solution of the Fokker-Planck equation (7) with velocity dependent diffusion coefficient
$D_{ql}(v)$
for the distribution function $f(v,v_0,t)$ with initial conditions\\
$f(v,t=0)\,=\,\delta (v-v_0)$ may be given in the form \be f(v,v_0,t)= C(t) \exp
{\left(-{Y^2(v,v_0)\over4t}\right)}, \ee with \be Y(v,v_0) =\int_{v_0}^v {du\over \sqrt {D(u)}}, \ee and $C(t)$
to be taken from the condition of normalization \be C^{-1}=\int {dv \exp {\left(-{Y^2(v,v_0)\over 4t}\right)}}.
\ee This approximation was proposed in Ref.\,4, however with other $C(t)$, which does not gives a proper time
scaling of dispersion.

For a time dependent diffusion coefficient the approximate WKB-type solution could be generalized as \be
f(v,v_0,t)= C(t) \exp {\left(-\frac{1}{4}Z^2(v,v_0,t)\right)}, \ee \be Z(v,v_0,t) =\int_{v_0}^v \frac{du}
{\sqrt{\int\limits_0^t D(u,\tau) d\tau}}. \ee Here, similarly to the previous case, $C(t)$ should be defined
from the condition of normalization.

\vspace{0.3cm}
\section{Comparison of simulation with numerical and analytical solutions of the Fokker-Planck equation}

In this section results of simulations are compared with numerical and analytical solutions of the Fokker-Planck
equation. For small Kubo numbers the solutions of Fokker-Planck equation with asymptotic quasilinear diffusion
coefficient give the similar evolution of velocity dispersion and average velocity as the solutions with time
dependent diffusion coefficient. In addition, for very small Kubo numbers at the beginning of evolution the
velocity dispersion grows almost linearly. Whether Kubo number increase the deviation from the linear law due to
dependence of diffusion coefficient on velocity becomes evident, however the solutions with $D_{ql}(v)$ and
$D(v,t)$ are still rather close. For a moderate Kubo numbers of the order of the unit, the difference between
solutions with time dependent and asymptotic diffusion coefficient becomes noticeable.

In Fig.\,2 the curves obtained in simulation for a wide spectrum and moderate field are compared with the
numerical solution of Fokker-Planck
equation and the WKB solution (17),\,(18).\\

\begin{figure}[h]
\includegraphics[width=7.5cm]{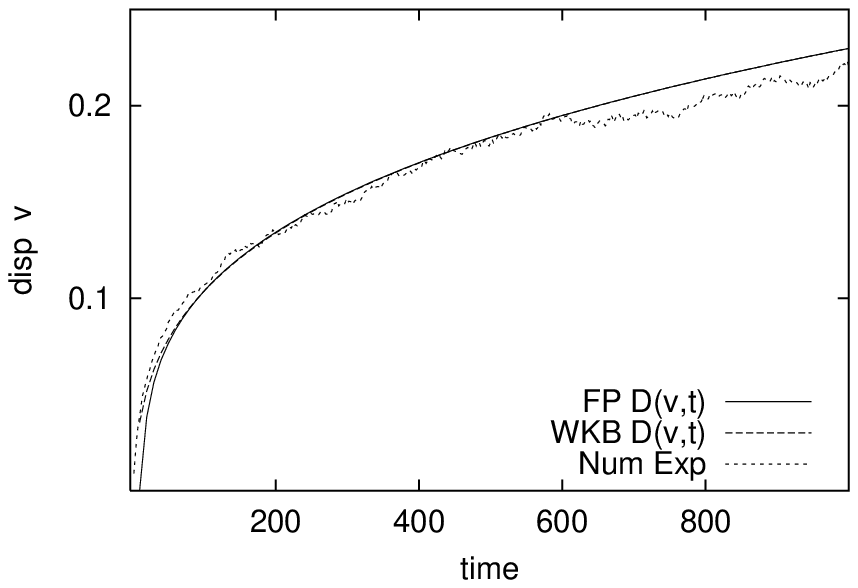}
\includegraphics[width=7.5cm]{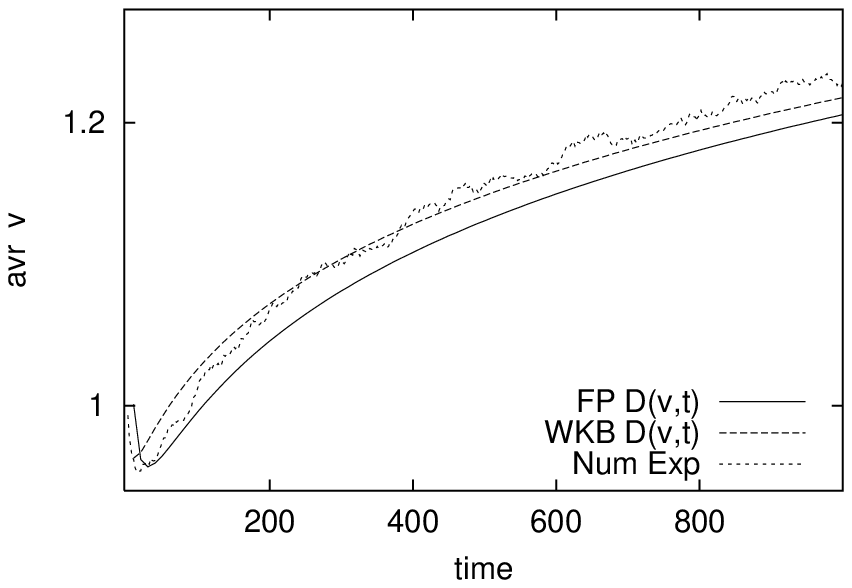}\\
\begin{center}
{\it Fig.\,2. Dispersion $\langle \Delta v^2 \rangle$ and average velocity $\bar{v}$ for a wide spectrum and
moderate field, $d=0.4$, $\sigma=0.01$ and $v_0=1$. Kubo number $Q=0.25$. Simulation and solution of
Fokker-Planck equation with diffusion coefficient $D(v,t)$ are compared with WKB solution (17),\,(18).}
\end{center}
\end {figure}

\noindent In Fig.\,3 is shown how WKB solution reproduces the early evolution of $\langle \Delta
v^2\rangle$ in the case of a narrow spectrum and moderate field with initial jump of dispersion.

\begin{figure}[h]
\begin{center}
\includegraphics[width=7.5cm]{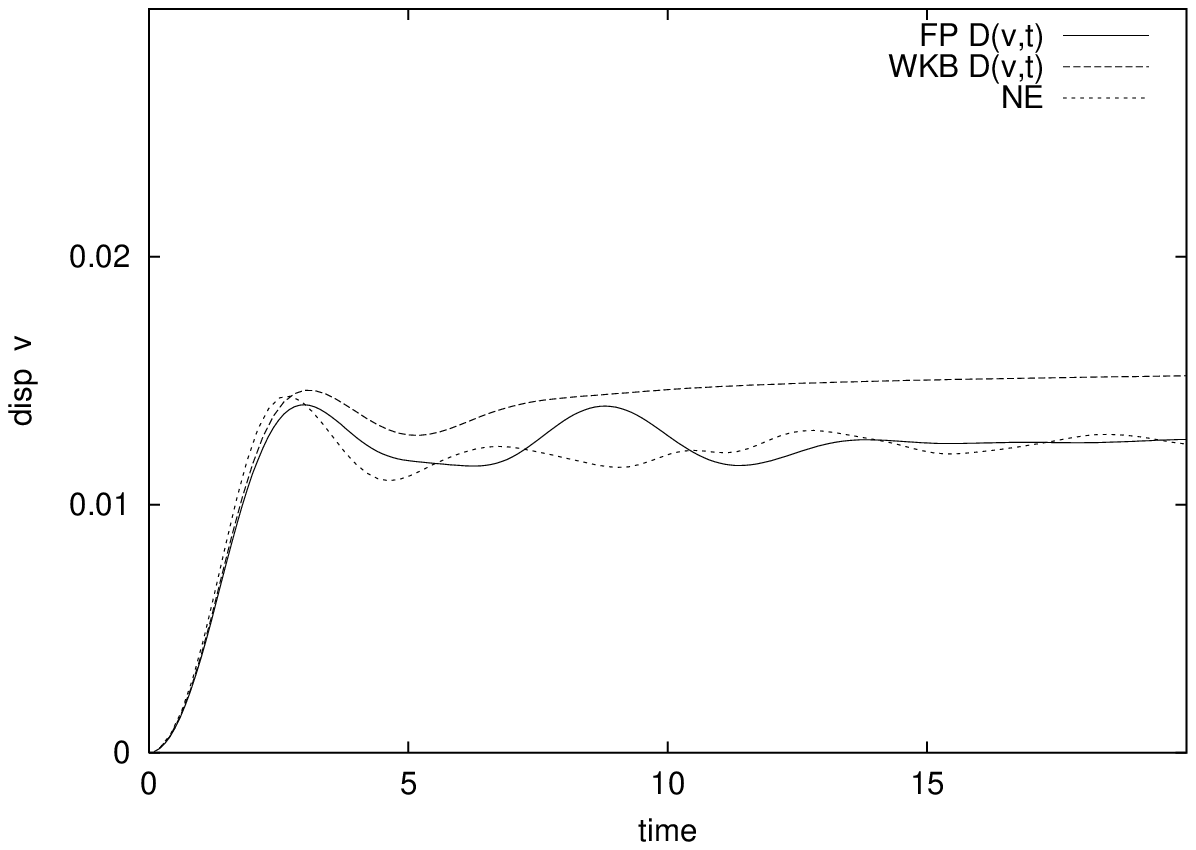}\\
{\it Fig.\,3. Dispersion $\langle \Delta v^2 \rangle$ for a narrow spectrum and moderate field, $d=0.04$,
$\sigma=0.01$, $v_0=1.0$. Kubo number $Q=2.5$. Simulation and solution of Fokker-Planck equation with $D(v,t)$
are compared with WKB solution (17),\,(18) in a small time scale.}
\end{center}
\end {figure}

In the cases of broad spectrum and low intensity particles for long time slowly diffuse on small distance on
$v$, which is less than halfwidth of $D_{ql}(v)$; and this time is enough for $D(v,t)$ to evolve to its
asymptotic value. In the opposite case, corresponding to Fig.\,3, particles on times substantially less then
$\tau_\cor$, while $D(v,t)$ is a broad in $v$ (c.f. Fig.\,1), diffuse at large distance which is more than
halfwidth of $D_{ql}(v)$.

The above examples were given for particles which initial velocity are not so far from the phase velocity of the
central harmonic. Such particles in each instant are in resonance with some harmonic of considerable intensity
and diffusion prevail over oscillations. For particles which initial velocities are far from resonance with
intensive harmonics the diffusivity is small and oscillations become more distinct. In Fig.\,4 dispersion are
given for nonresonant particle. The curves obtained in simulation, as numerical solution of the Fokker-Planck
equation with time dependent diffusion coefficient and in WKB approximation are shown.

\begin{figure}[h]
\begin{center}
\includegraphics[width=7.5cm]{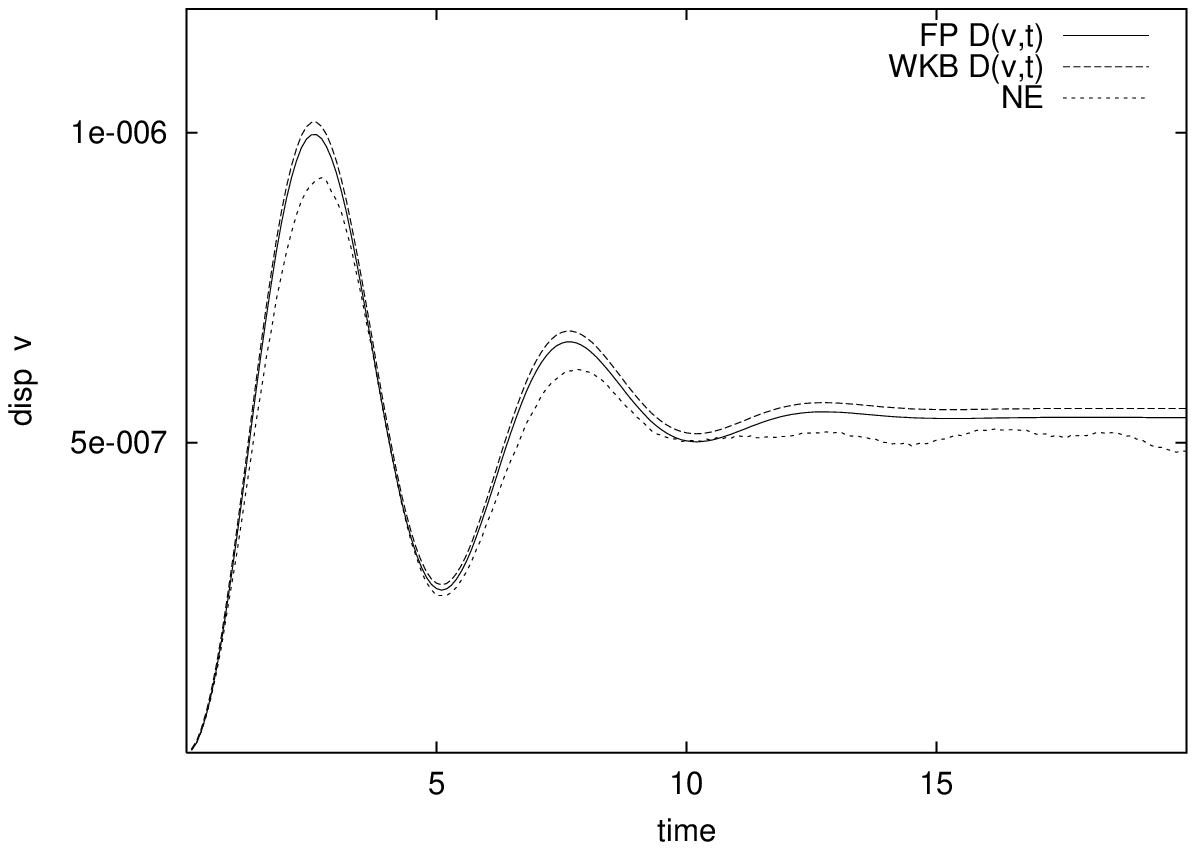}\\
{\it Fig.\,4. Dispersion of nonresonant particles $\langle \Delta v^2 \rangle$ for a narrow spectrum and weak
field, $d=0.04$, $\sigma=0.0001$, $v_0=1.2$. Kubo number $Q=0.25$. Simulation and solution of Fokker-Planck
equation with diffusion coefficient $D(v,t)$ is compared with WKB solution (17),\,(18) on a small time scales.}
\end{center}
\end {figure}

\vspace{0.3cm} \noindent
\section{Power law dispersion}
Simulation shows the velocity dispersion obeys a power law dependence on time.  Numerical solutions of
Fokker-Planck equation give the same power law, as simulation, and it easily could be calculated for much longer
time. Such power law is also recovered from WKB approximation, and in this case it could be related to power law
dependence of normalizing constants $C(t)$. The example with $Q=0.079$ is given in Ref.\,7. Here, in Fig.\,5 the
plot is given for the different Kubo number, $Q=0.25$, for numerical and WKB solutions of Fokker-Planck equation
along with time dependence of $G=C(t)\,t^q$ (note that $q\,=\,p/2$).

\begin{figure}[h]
\includegraphics[width=7.5cm]{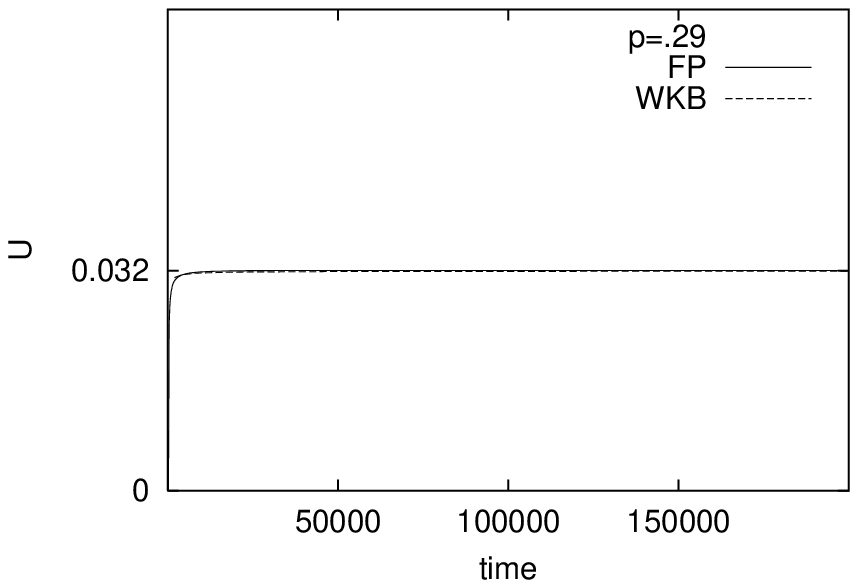}
\includegraphics[width=7.5cm]{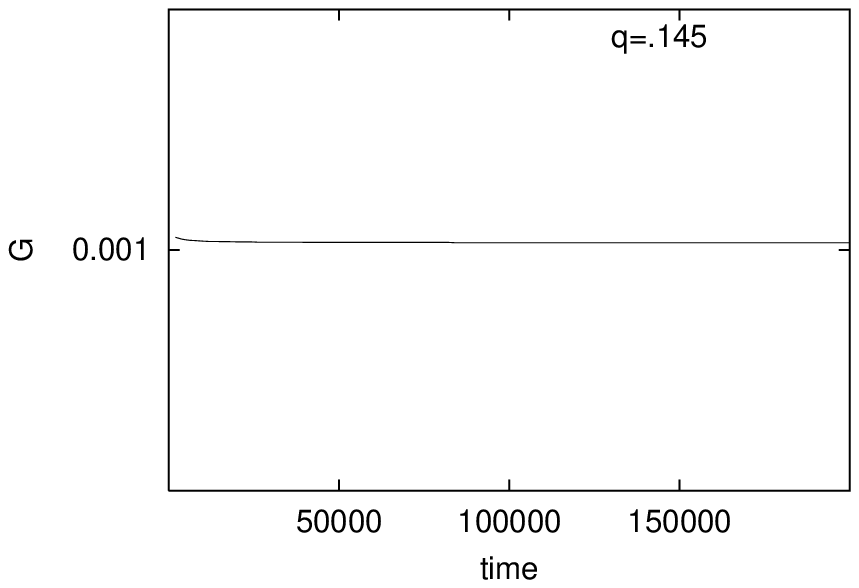}\\
\begin{center}
{\it Fig.\,5. $U={\langle \Delta v^2 \rangle/t^p}$ against $t$. Power law dispersion for a wide spectrum,
$p=0.29$ (left), and $G=C(t)\,t^q$ against $t$, $q\,=\,p/2\,=\,0.145$ (right);  $d=0.4$, $\sigma=0.01$, $v_0=1$,
Kubo number $Q=0.25$.}
\end{center}
\end {figure}

\vspace{0.3cm} \noindent The velocity dispersion shows power law time dependence, with exponent dependent on
particular spectrum. In terms of WKB solution it is related to power law dependence of normalizing constant
$C(t)$.

\vspace{0.3cm}
\noindent
\section*{Conclusions}
For small field intensity and wide spectrum (Kubo number less than the unit)
the solution of the Fokker-Planck equation
with quasilinear diffusion coefficient
gives good agreement with results of numerical experiment. To have
a consistency for high intensity and/or narrow spectrum
(Kubo number of the order or larger than the unit)
the generalization of the
Fokker-Planck equation is to be done by introducing a time dependent
diffusion coefficient. An analytical approximation for such
solutions is proposed. Velocity dispersion manifests power law
time dependence and the exponent is dependent on the spectrum.

\vspace{0.2cm}
\noindent
{\bf Acknowledgments}.
Two of the authors (V.Z. and A.Z.) are grateful
to Chalmers University of Technology for their hospitality.

\vspace{0.4cm}
\noindent
$[1]$ I~Doxas, J~R~Cary {\it Phys. Plasmas}  {\bf 4} 2508 (1997)\\
$[2]$ J~R~Cary, I~Doxas, D~F~Escande,A~D~Verga {\it Phys. Fluids}
 {\bf 4} 2062 (1992) \\
$[3]$  A~S~Bakaj {\it Dokl. Acad. Nauk SSR} {\bf 237}, 1069 (1977);
A~S~Bakaj, Y~S~Sigov {\it ibid}, 1326\\
$[4]$ E~Vanden Eijnden  {\it Phys. Plasmas}  {\bf 4} 1486 (1997)\\
$[5]$ F~Doveil, D~Gresillon {\it Phys. Fluids}  {\bf 25} 1396 (1982)\\
$[6]$ S~A~Orszag, R~H~Kraichnan. {\it Phys. Fluids}  {\bf 10} 1720 (1967)\\
$[7]$ A~Zagorodny, V~Zasenko, J~Weiland
      {\it 23th EPS Conference on Contr. Fusion and Plasma Phys., St. Petersburg,
      7-11 July 2003 ECA} Vol. 27A, P-2.3\\

\end{document}